# Recent Developments in Detection of Central Serous Retinopathy through Imaging and Artificial Intelligence Techniques – A Review

**Syed Ale Hassan[1], Shahzad Akbar[1], Amjad Rehman[2], Tanzila Saba[2], Hoshang Kolivand[3], Saeed Ali Bahaj[4]**

[1] Riphah College of Computing, Riphah International University, Faisalabad Campus, Pakistan
[2] Artificial Intelligence & Data Analytics Lab (AIDA), Prince Sultan University, Riyadh, Saudi Arabia.
[3] School of Computer Science and Mathematics, Liverpool John Moores University, Liverpool, UK
[3] School of Computing and Digital Technologies, Staffordshire University, Staffordshire, UK
[4] MIS Department College of Business Administration, Prince Sattam bin Abdulaziz University, Alkharj, Saudi Arabia.

Corresponding authors: Shahzad Akbar (shahzadakbarbzu@gmail.com) and Amjad Rehman (rkamjad@gmail.com)

**ABSTRACT** Central Serous Retinopathy (CSR) or Central Serous Chorioretinopathy (CSC) is a significant disease that causes blindness and vision loss among millions of people worldwide. It transpires as a result of accumulation of watery fluids behind the retina. Therefore, detection of CSR at early stages allows preventive measures to avert any impairment to the human eye. Traditionally, several manual methods for detecting CSR have been developed in the past; however, they have shown to be imprecise and unreliable. Consequently, Artificial Intelligence (AI) services in the medical field, including automated CSR detection, are now possible to detect and cure this disease. This review assessed a variety of innovative technologies and researches that contribute to the automatic detection of CSR. In this review, various CSR disease detection techniques, broadly classified into two categories: a) CSR detection based on classical imaging technologies, and b) CSR detection based on Machine/Deep Learning methods, have been reviewed after an elaborated evaluation of 29 different relevant articles. Additionally, it also goes over the advantages, drawbacks and limitations of a variety of traditional imaging techniques, such as Optical Coherence Tomography Angiography (OCTA), Fundus Imaging and more recent approaches that utilize Artificial Intelligence techniques. Finally, it is concluded that the most recent Deep Learning (DL) classifiers deliver accurate, fast, and reliable CSR detection. However, more research needs to be conducted on publicly available datasets to improve computation complexity for the reliable detection and diagnosis of CSR disease.

## I. INTRODUCTION

The retina is located behind the eyeball near the optic nerve and comprises a thin layer of tissue [1]. It obtains the focused light from the eye-lens, converts it into neural signals, and imparts signs to the brain for visual recognition. The retina processes light using a layer of photoreceptor cells. These are light-sensitive cells responsible for detecting visual characteristics, such as color and light intensity. Subsequently, the data accumulated by the photoreceptor cells are sent to the brain through the optic nerve for optical recognition. Therefore, the retina plays a crucial role in image processing for the human brain recognizes and distinguishes various surrounding objects and names them.

Since any damage to the retina may have severe ramifications to our ocular abilities. A typical schematic diagram of the human eye has been depicted in Figure 1.

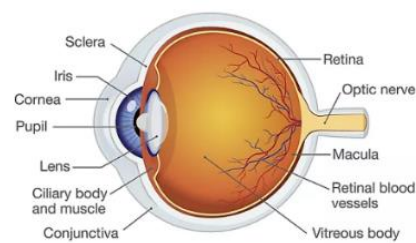

**FIGURE 1: Anatomy of the Human Eye, depicting various parts including the Retina** [2]

In this review, various retinal disease diagnosis and detection techniques have been described, and then focused on utilizing advanced imaging and Artificial Intelligence (AI) methods in a specific retinal disorder known as the Central Serous Retinopathy (CSR).

The retinal disease diagnosis and detection can be performed through several imaging techniques such as fundus photography and Auto-Fluorescence, Optical Coherence Tomography (OCT), and Angiography. Recently, in advanced Deep Learning (DL) investigations, every patient is potentially a source of precious diagnostic data that can be utilized in training new Machine Learning (ML) models for better treatment and diagnosis [3]–[5]. The most recognized use of artificial intelligence (AI) strategies in retinal disease are the development of spots intricate to disease characteristics on color fundus photos. In this context, programmed segmentation and characterization of anatomic and pathologic have been combined and applied in retinal diseases [6], [7]. These techniques are effectively being utilized due to the emergence and popularity of the OCT methodology since the early 2000s.

Consequently, it is now possible to conduct exact detection of several features to perform automated segmentation of retinal diseases such as, sub-retinal and intraretinal fluid, drusen, color epithelial disintegration, and topographical atrophy with equivalent exactness as human graders. Another research field is developing a profound-settled image of the retinal vasculature, where OCTA generates multiple consecutively generated images using OCT-B scans. In recent times some automated systems have been designed in the detection and treatment of retinal disease mentioned as Diabetic Retinopathy [8]–[14], Hypertensive Retinopathy [15]–[18], Age-related Macular Degeneration [19]–[23] and Glaucoma [24]–[29].

The automatic diagnosis of retinal disorders through the analysis of retinal images has become a significant practice in clinical systems. Using these automated techniques, the physicians have extra optimized and accurate results. In the past, manual methods were used for retinal disease detection, which were inefficient, time-consuming, and inaccurate. Contrarily, computer-aided retinal disease recognition systems are cost-effective, objective, accurate, user-friendly, and fast. In addition, they do not largely rely on the ability of an experienced ophthalmologist to examine the various scanning images to detect the disease [29]. This review article focuses on a specific retinal disease that is commonly known as Central Serous Retinopathy (CSR) or Central Serous Chorioretinopathy (CSC). It was first reported by Von Graefe as "*Relapsing central Lueitic Retinitits*" [30]. The CSR is among one of the major eye diseases, which is caused due to the collection of fluids behind the retina, which can severely damage eyesight due to the presence of a delicate tissue layer constituting retina. Therefore, early-stage detection can essentially allow for taking the appropriate preventive measures to restore vision, thereby leading to complete recovery. The statistics have shown that CSR usually affects one eye, but the damage to both eyes cannot be ruled out during the lifetime of any patient. Besides, in some cases, the patients may recover after a while, without any treatment [31]. Moreover, CSR usually is not correlated with affecting patient's quality of life [32]. According to many studies, the CSR patients' ages range between 7 to 83 years, with the most affected age group of 40-50 years.

The interpretation of CSR in classical terms is referred to as acute CSR. A patient with intense CSR may encounter obscured vision, decrease in contrast sensitivity and shading vision, metamorphopsia, and minor hyperopic move. Traditionally there is a central serous separation of focal retina, sometimes with dull yellow stores and in a few cases with serous RPE separation. In several cases of CSR, a permanent sub-retinal fluid accumulates for three months or more: resulting in permanent visual symptoms. Such cases of CSR frequently experience a fluctuating degree of sub-retinal fluid. However, most cases recover automatically, whereas some patients may experience chronic CSR [33].

Nevertheless, their visual acuity remains typically steady. The chronicity of these patients depends upon the time duration of CSR, and it usually takes 3 to 6 months in case of acute CSR. Contrarily, in chronic CSR patients, the symptoms of morphological changes and an increased risk of CNV have been observed [33].

Patients with CSC are generally of the age of 25 and 50, in which men are afflicted far more frequently than women. These patients normally have symptoms such as the grumblings of unexpected beginning, contortion, and focal vision blurring. Visual acuity ranges between 6/5 and 6/60, but its usual range is 6/9 to 6/12 [34]. In the case of CSR, it is generally a self-constraining disease with unconstrained resolution having boundaries of 3–4 months. Historically data reveals that nearly half of the CSR patients may experience recurrences of the disease within a year, causing the patient to undergo various treatment procedures, which may last for three months in chronic CSR, recurrent CSR, and first-time CSR patients. Some standard CSR treatments include the Micro Pulse Laser Treatment (MPLT), the Transpupillary Thermo treatment (TTT), the Photodynamic therapy (PDT), and the Intravitreal anti-Vascular Endothelial Growth Factor (anti-VEGF). These treatment methods are based on the following points:

i. A majority of the population automatically recuperates within 4 to 6 months without requiring any specific medicaments.
ii. If a patient's CSR lasts for a year, then some treatment may be required.
iii. In rare cases, if CSR lasts for more than a year, an ophthalmologist may opt for specific treatments such as RPE detachment or bullous retinal detachment [35].

There have been approximately 40 articles published related to deep learning and medical imaging. Among these 40 articles, only 29 articles are associated with Central Serous Retinopathy (CSR). These articles, along with their methodologies, advantages, and limitations, are discussed in this review. As mentioned above, CSR detection is performed through various imaging technologies. We briefly summarize them below.

### A. Imaging Technologies for Detection of CSR

The traditional imaging technologies for CSR detection include Fundus Photography, Fluorescein Angiography (FA), and Optical Coherence Tomography (OCT). These technologies are discussed in the following subsections.

### 1) THE FUNDUS PHOTOGRAPHY

Fundus Photography is a process of obtaining the retinal red-free image and is considered an alternative to OCT imaging. This technique is based on the statistical approach and requires advancement in contrast to its forerunner, color photography film. Similarly, the digital image of retina provides quick towering-resolution and consistent image, and it is accessible instantly and manageable for the development of an image. Moreover, Fundus photography is regularly employed for ailment records and clinical examination, along with potential usage for tolerant training and telehealth. Additionally, the images generated through Fundus techniques can incorporate average and extensive views [3]. Figures 2 and 3 depict retinal scans obtained via fundus photography [36], [37]. In Figure 2, the normal indications of a healthy eye have been depicted, whereas, in Figure 3, the dark spot of a blister of fluid caused by CSR disease has been shown.

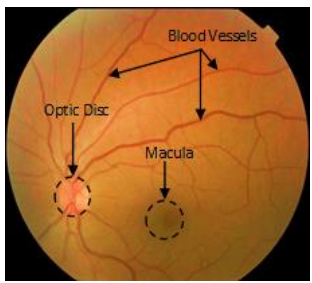

**FIGURE 2: Normal Retinal Fundus scan**

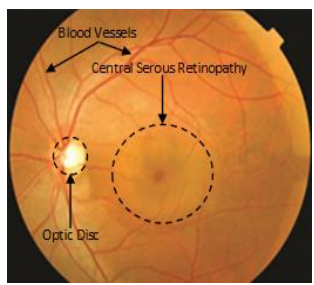

**FIGURE 3: Retinal Fundus scan with CSR**

In CSC, multimodal-imaging methods incorporate OCT with improved deep imaging, Indocyanine green angiography, fundus auto fluorescence, OCTA, and fluorescein angiography. , The advancement of recent scanning strategies, has transformed the comprehension regarding the pathophysiology of CSC, and thus the treatment has thoroughly changed. This review paper explains existing comprehension about physiopathology and hazardous components and multimodal imaging-based highlights of CSR.

### 2) FLUORESCEIN ANGIOGRAPHY (FA)

Fluorescein Angiography (FA) is an effective imaging technique in which fluorescent dye is injected into blood vessels of patient's eyes in order to capture their clear images. The main objective of this technique is to highlight the blood vessels to form a clear and visible image. The patient is normally prescribed with primary care before initiating the FA procedure to ensure a satisfactory blood stream in the veins. In addition, the physicians propose primary care to analyze further the eye issues, including macular degeneration, diabetic retinopathy, or Central Serous Retinopathy (CSR). The retinal images before and after fluorescein angiograms are depicted in Figures 4 and 5, respectively [36].

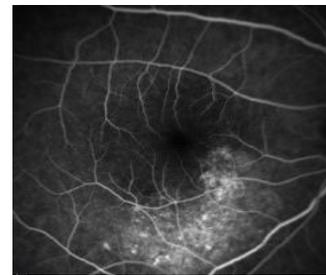

**FIGURE 4: Retinal image before the Fluorescein Angiograms (FA)**

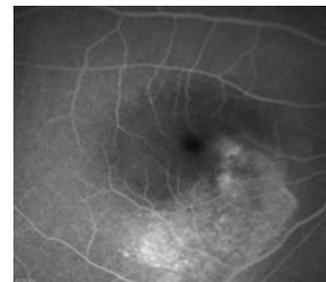

**FIGURE 5: Retinal image after the Fluorescein Angiograms (FA)**

### 3) OPTICAL COHERENCE TOMOGRAPHY (OCT)

Optical Coherence Tomography (OCT) appeared as an advanced automation technology used for detection and diagnosis of different diseases developed in the early 1990s. Above mentioned approach is similar to ultrasound imaging. However, instead of making use of sound it uses the light. The combination of catheters and endoscopes with OCT produces high-resolution imaging of the organ system. OCT can provide tissue images intangible and situ form. Additionally, two essential modalities of OCT are named as (i) Time-Domain (TD), and (ii) Spectral Domain (SD) [37]. Time domain is used to detect the interference patterns in creating a picture, whereas the Spectral Domain SD fuses a spectrometer to build examining time. SD can get pictures more rapidly than TD due to its effective imaging method.

The physiologic segment gets the visual light beam partitioned into the pigmented part (pigment epithelium) and the nervous part. The layers of retina are named as Ganglionic cell layer, inner plexiform layer, Nerve fiber layer, Outer plexiform layer, External limiting membrane, Internal limiting membrane, Outer nuclear layer, Retinal pigment epithelium, Inner nuclear layer, Photoreceptor outer segment, Interface between IS and OS, Photoreceptor inner segment and Outer Photoreceptors[38]. These layers are present in retina OCT as shown in Figure 6.

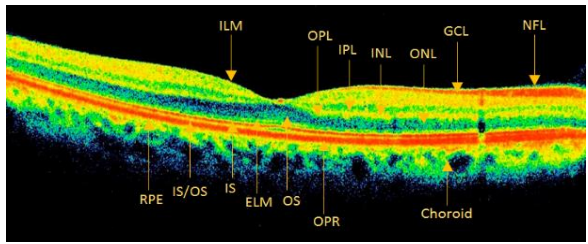

**FIGURE 6:** Normal retinal OCT image [33] RPE: Retinal Pigment Epithelium, IS/OS: Interface between IS and OS, IS Photoreceptor Inner Segment, ILM: Inner Limiting Membrane, ELM: External Limiting Membrane, OS: Photoreceptor Outer Segment, OPL: Outer Plexiform Layer, IPL: Inner Plexiform Layer, OPR: Outer Photoreceptor/RPE complex, INL: Inner Nuclear Layer, ONL: Outer Nuclear Layer, GCL: Ganglionic Cell Layer, NFL: Nerve Fiber Layer.

In Figure 6, the layers of the OCT image of retina do not indicate any abnormality or disease [39]. Normally a person with perfect vision has the same OCT as depicted in Figure 6. In Figure 7, the sub retinal fluid and intra-retinal fluid (OCT Scan) are linked with the CSR disease [40]. Besides, the patient suffering from CSR produces a different OCT scan, as depicted in Figure 7.

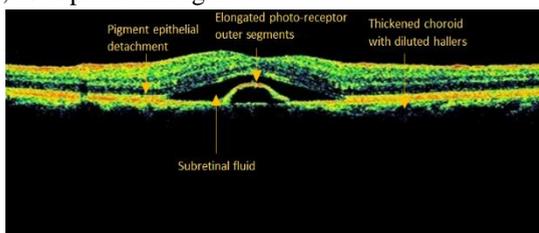

**FIGURE 7:** Retinal OCT image with CSR disease

### B. OCT Imaging Datasets
In this study, two publicly available datasets associated with CSR are analyzed. A typical dataset contains a range of records, and these CSR datasets consist of the repositories of OCT and fundus images. Several researchers normally access and utilize these publicly accessible datasets, which can be easily retrieved using their specific links. In various experimental studies, a portion of the all-out images is normally downloaded from these datasets in order to train new Machine Learning models and algorithms and achieve testing goals. In the following list, we briefly describe two publicly-available datasets along with their accessibility.

#### 1) OCTID DATASET
The OCT imaging database is an open-source OCT imaging database, which is accessible at the website of University of Waterloo, Canada. This database contains 102 MH, 55 AMD, 107 DR, as well as 206 NO retinal images. The interface of the dataset is user-friendly, and all the diverse groups have been classified in the form of discrete datasets. The users can easily access the required datasets via a unique DOI link. Additionally, the images can be observed and downloaded in the form of folders or as zip files [41].

#### 2) ZHANG'S LAB
Zhang's lab [42] contains the largest dataset of labeled OCT images. There are thousands of publicly available datasets of OCT images. It has been referred to specifically in identifying medical diagnosis and treatable disease by image-based deep learning. At Zhang's lab, various images are available for both training and testing purposes of various Machine Learning ventures in retinal disease detection.

This review article provides a comprehensive review of several state-of-the-art technologies, which make use of Artificial Intelligence (AI) algorithms for automatic detection of CSR disease in patients. Most of these algorithms lead to the evolution of new models that are trained on proprietary datasets that are also discussed in this study. These models have assisted in development of many commercially-available products used by hospitals and ophthalmologists to automatically detect CSR in patients. To the best of our knowledge, no such study is available in literature for summarization of such technologies for CSR disease detection.

The remainder of this article has been organized as follows: Section II describes a detailed review of the literature. It also contains the methodology to organize this review article by discussing the content inclusion and exclusion criteria. Furthermore, it includes case studies as well as Artificial Intelligence (AI) techniques in imaging, using machine learning and deep learning. Section III contains an analytical discussion based on the literature review and establishes certain research findings and open questions for researchers. Finally, Section IV provides a detailed conclusion.

## II. LITERATURE REVIEW
In this section, a comprehensive literature review has been compiled based on the following steps:

    i.   Identification of relevant and advanced research articles.

    ii.   Articles emphasizing on the basic understanding the research problem.

    iii.   A comprehensive search strategy that identifies the research problem.

iv. Extraction of the desired data from the selected research articles.

v. Validation and fact-checking of the collected data.

vi. Representation of data for better visualization and readability.

The literature review articles have been searched from the following reputable and well-recognized scientific publishers and organizations:

i. The American Academy of Ophthalmology, (http://www.aaojournal.org/)

ii. The JAMA Network, (http://jamanetwork.com/journals/jama)

iii. Investigative Ophthalmology and Visual Science, (http://iovs.arvojournals.org/)

iv. American journal of ophthalmology, (https://www.ajo.com/)

v. Digital library of IEEE Xplore, (http://ieeexplore.ieee.org)

vi. Springer Nature Link, (http://link.springer.com/)

vii. Elsevier, (https://www.elsevier.com/)

viii. Science Direct, (http://www.sciencedirect.com/)

These scientific databases have been accessed to accumulate peer-reviewed journal articles, review articles, and conference papers. Figure 8 illustrates the methodological flowchart of this review with different phases.

In Figure 8 the methodological phases through which this review article has been arranged and presented for the researchers, scientists, physicians, and practitioners to get CSR treatment insights through detection using artificial intelligence techniques. There are four different phases of the review process, as shown in Figure 8.

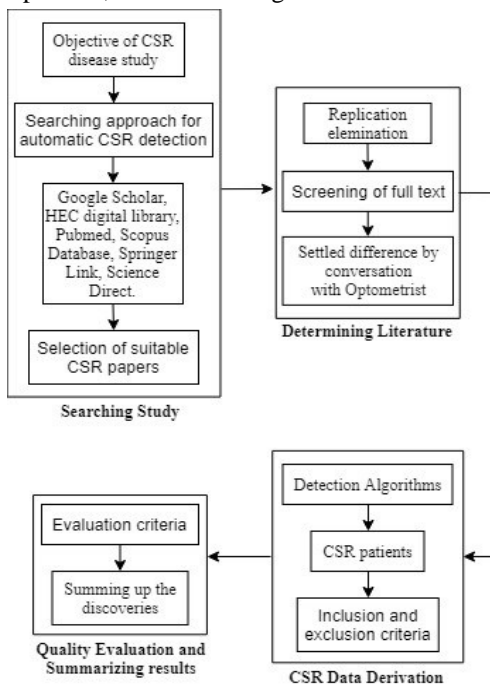

**FIGURE 8:** Flow chart of systematic review process

## 1) SEARCHING STUDY

In this phase, the objective of CSR disease study is defined and elaborated. Based on a defined selection criterion for automatic CSR detection, a number of suitable CSR papers have been selected for further analysis.

## 2) DETERMINING LITERATURE

After an extensive search of relevant articles, the next phase was to determine the valuable articles for this study. For that purpose, all the repetitive articles were eliminated based on their citations, authors' integrity, and institutional affiliations. After studying the text of each article, the eye physicians were consulted for settling differences and a better understanding of the CSR disease.

## 3) CSR DATA DERIVATION

Various advanced CSR detection algorithms based on AI, ML, and DL were studied and analyzed in this phase. This process provided a list of state-of-the-art algorithms based on inclusion and exclusion criteria.

## 4) QUALITY EVALUATION & RESULTS SUMMARIZATION

Finally, all the discoveries have been summarized after evaluating the quality of the content.

In the following subsections, an extensive review of various technologies based on advanced research articles is provided. First, the content inclusion/exclusion criteria are detailed. Then, the criterion of journal selection is briefly described, which is followed by goal of this review.

## 5) CONTENT INCLUSION CRITERIA

"*Detection of CSR in OCT images using artificial intelligence*" is set as the sign indication for literature search. The abstracts, titles, and methodologies of each article have been checked for the research prompt and facts. All those articles were selected which have a clear research domain of identification and detection of CSR through retinal images. Another criterion was set to check the use of data sciences and Artificial Intelligence in CSR disease detection.

## 6) CONTENT EXCLUSION CRITERIA

The aim of this review is to study and organize algorithms and techniques that have been used for the segmentation of CSR from retinal images. In this way, the articles having algorithms but not relevant to CSR have not been considered in this review. Additionally, the articles were selected based on the repute and the impact fact of the journal. In the case of conference/symposium papers, their proceedings and ranking of the conference were considered for this review. Another criterion was to weigh the articles on the basis of their citations and institutional affiliations. Finally, the contents of similar papers have not been considered to maintain the integrity in the quality of the review article.

## 7) SELECTION OF JOURNAL ARTICLES

The research articles were selected by cautious investigation for its consideration as well as prohibition standards. The selected articles were copied in the system library and spared, the terminology used to spare articles represent title, publishing year, publisher, author first and last initials. Such description facilitated in arranging the research articles and their proficient recovery as per inquiries. References of the considerable number of papers have been taken and spared in *EndNote* reference library. Based on this selection criterion, a database of articles was set up which included articles related to basic understanding of retinal disease based on studies of human anatomy, the CSR disease and its variants, and the recent studies of utilization of AI techniques for automatic detection of CSR. In the following sections, these studies are discussed in detail.

8)  GOALS OF THIS REVIEW

In this review, the sequel of precise and complete study of CSR algorithms has been explored for automatically segmenting process in retinal OCT and fundus images. The significant goal of the review is to analyze the development of CSR algorithms and recognition of CSR from fundus images. The analysis of optimistic algorithms has some applicable clinical attributes for diagnosis, treatment, and prevention. This review article presents the novel insights in using AI techniques for CSR. As far as we could possibly know, there is no such review in literature to give a thorough review of such technologies. The significance of this review is to present recent pattern recognition technologies, based on the methodologies in pre-processing and segmentation from OCT and fundus images that have already detected the CSR disease.

9)  DETECTION OF CSR BASED ON IMAGING TECHNOLOGIES

In this section, several imaging technologies for CSR detection are discussed. In each case, the relevant research studies are pointed out, along with their brief descriptions and possible shortcomings.

M. M. Teussink et al. [43] had differentiated OCT angiography features of a patient with acute CSR considering FA as well as Indo Cyanine Green Angiography (ICGA). In conducting experiment, the research team included 225 eye images along with 100 FA and ICGA images. Additionally, 125 and 30 OCT angiography images of patients were recorded. The results of chronic CSR reflected the irregular choriocapillaris flow patterns. Several researchers have studied the combination of both FA and ICGA [44], [45]. However, this study has been confined to a modest number of patients. Further improvements in knowledge regarding vascular changes require prospective and longitudinal studies.

J. Y. Shin Velthoven et al. [46] had used ophthalmoscope to detect CSR from OCT scans**.** The CSR detection was normally depending on FA or biomicroscopy. The sample data of scanned eye images of 38 patients for testing purposes was considered. The OCT ophthalmoscope usually provides additional knowledge related to the zone involved and is practically identical to the statistics proposed for FA. Therefore, Ophthalmoscope proves to be the best technique to detect CSR at an early stage. Two recent studies confirmed the findings of different PEDs on OCT in patients with CSR. Hence, FA has additionally exhibited zones of diffuse epitheliopathy in patients suffering from constant CSR [47].

Furthermore, E.Costanzo et al. [48] had compared the OCTA with multimodal imaging, which resulted in CSR detection. The authors have performed experiments of both techniques using the same dataset of 33 eyes of 32 patients. The population and protocols in this study have been analyzed to find the results of both techniques. Consequently, the resultant dataset of both techniques has been naive compared to the previous studies in the treatment of eye diseases. This procedure used the description and classification of abnormal choroidal vessels. The experimental analysis declares OCTA had all the earmarks of being a promising method since it kept away from the weight of intravenous injections of dye which probably would complicate genuine symptoms. The findings of this study verified that OCTA technique was eligible for the detection of an unusual choroidal vessel specimen among all of the cases. Similar studies using the OCTA method had the option to identify the distinctive neovascular network, usually CNV, in 58% of eyes with lingering CSR [49]**.** Another similar study resulted in excessive specificity and sensitivity of 100% in OCTA images [50].

F. Rupesh Agrawal et al. [51] proposed early detection of CSR with the help of an image binarization. For this purpose, a sample of 78 eyes from 39 patients was extracted for the experiment. The significant vascularity records with intense CSR compared and clear solid fellow eyes were observed. The choroidal vascular changes in CSR of [52], [53] had resulted in similar observations. The CVI was observed critical, and LA was lower in the fellow eyes compared to the age-coordinated healthy subjects. The stromal zone had not been significant for exactness as the age-coordinated selection of healthy subjects was required. Subsequently, the general vascularity index was present in the fellow eyes compared to the age-matched healthy eyes that generally make the eyes prone to leakage. The authors also stated that they could not use normalization before binarization.

S. Weng Xiang et al. [54] proposed the OCTA to tackle CSR disease. This research was conducted on seventy patients and seventy-five eyes were utilized to perform OCTA on them. Two readers analyzed the obtained images and CNV was evaluated at outer retina. The authors concluded that the CSR disease (along with the CNV disease) mostly occurs in patients over the age of 50 years [55]**.** The authors collected samples of 70 eyes of patients as dataset for their experiments. The advanced mediums for photography were used which had fundus FA. To detect CSR in images, the OCTA appeared to be the best algorithm and stated that no other technique could perform the detection similar to this technique. The results of the experiments represent 97.77% accuracy, approximately 100% sensitivity, and a specificity level of 93.33%. The authors have compared their technique with those of Quaranta-El Maftouhi et al. [49], who claimed that 12 eyes suffering from chronic CSC displayed the characteristic of the choroidal hyper permeability except authentication of CNV.

Michael R et al. [56] proposed a rational technique for OCT image scan. This technique develops high resolution tomography for ocular tissues. The sample dataset of 16 patients for examination purpose was taken and results were correlated with slit-lamp biomicroscopy and fundus photography [57]. Likewise, optical imaging tomography has provided a quantitative assessment for the accumulation of subcutaneous fluid, which can be used to achieve irregular goals of solid separation with high efficiency [58].

In addition to the above research, Z. W. Jelena Novosel et al. [54] proposed the locally-adaptive and loosely-coupled approach to detect retinal fluids in CSR disease. The dataset from an in-process study on CSR in a hospital in Germany was collected. The results were compared by medical doctors manually and by using automatic trained segmentation. Furthermore, the authors compared their algorithm with other similar researches in a subjective manner. A few methodologies for the division of fluid-associated abnormalities reported the TPR ranging between 86 to 96%, [59], [60] that is similar to the authors attained 95% TPR. Overall, their algorithm presented a flexible solution for retinal fluid segmentation owning high accuracy qualitatively.

M.Wu Xiang et al. [61] proposed the max-flow optimization algorithm based on a three-dimensional (3-D) continuous method. This method covered two of the Neurosensory Retinal Detachment (NRD), together with the Pigment Epithelial Detachment (PED). The proposed method is based on a probability map using a random forecast model. The authors used 37 retinal OCT images for training and testing purposes. The experimental results have shown that 92.1%, 0.53%, 94.7%, and 93.3% constitute the Dice Similarity Coefficient (DSC), for segmentation with NRD as compared to 92.5%, 0.14%, 80.9%, and 84.6%, with PED segmentation. Various methods were proposed to conduct the classification of CSR but all of them had been observed as 2-D images which classified retinal disorder [62] with a few restrictions that need to be enhanced in future investigations. There were extremely little fluid sectors used as a rarity in the proposed methodology for classification of retinal fluids.

H. J. Chien et al. [63] proposed the Balloon Snake Algorithm implemented in MATLAB tool. This algorithm was utilized for the segmentation of fluid from OCT retinal images in terms of ophthalmology [64]. The primary objective of this research work was to detect edges and volume of retinal fluid and provide the best possible OCT images. The following steps have been referred for CSR disease detection.

   i.   Patient selection and data collection.

   ii.   OCT scan protocol.

   iii.   Sub-retinal fluid volume measurement via manual segmentation.

   iv.   Obtaining sub-retinal fluid contour and volume from the Balloon Snake algorithm.

The experimental results using the Balloon Snake algorithm were reported to be approximately 30 minutes faster as compared to the manual detection methods. The researchers have introduced different strategies for measuring the volume of retinal fluid such as manual, semi-manual technique or automatic commercially-available software products [65]. These techniques have been precise and regeneratable that may require human collaborations to develop the segmentation line. The Balloon Snake-based algorithm is capable of measuring a retinal fluid volume within 5 minutes. Using this methodology in clinical examinations is found to be extremely reliable, accurate, and fast.

K. Gao, et al. [37] had proposed instinctive technique which depends on graph theory and preceding B-scan information to automatically segment retinal layers in CSR diseases. The segmentation process includes the estimation of the external nuclear layer boundaries, inner limiting membrane, photoreceptor inner segments, abrasion of eye, and pigmented layer of retina and RPE-choroid on graph search model. The region of flexible search was developed with calculation of the thickness. The experimental results had shown that it meant entire thickness in contrast with manual segmentation of (3.68±2.96 μm) as well as (5.84±4.78 μm) respectively. Recent researchers also worked on automatic methods, but these strategies were dependent upon intense irregularities and breaks in the retinal layers because of outlines brought about by retinal veins. Future direction has been proposed to classify the border of CSC zone, which constitutes of different layers to measure CSC locale also breakdown variety of every retinal layer in various infections.

Similarly, SG Odaibo et al. [66] had proposed an optimized technique to detect CSR in OCT images through cloud-based mobile Artificial Intelligence (AI) platform. This technique played a vital role in assisting physicians with less cost and high-accuracy results. The algorithm is capable to classify fluid and non-fluid OCT images. Interestingly, this algorithm was integrated as an application in IOS operating system for a test run. After testing and experimenting, the authors reported the sensitivity levels between 82.5% & 97%, whereas the specificity range was reported between 52% & 100%. Moreover, the authors proposed to test large amount of dataset in their future experiments [56].

M. LU HE et al. [67] derived comparison of the multicolored imaging (MC) with the Color Fundus Photography (CFP) and employed the result achiever technique to detect retinal fluid in CSR. The authors took the sample of 75 eye scans of 69 patients for experimental setup. The examination of all the samples for the clarity of vision, intraocular pressure, slit lamp, CFP (Digital Fundus Camera, VISUCAM 200; Carl Zeiss Meditec AG, Jena, Germany), FFA, MC image, as well as SD-OCT was conducted. The comparison of results indicated that MC plays a more influential role of detection compared to the CFP. The MC method attained 92.0% detection rate compared to the CFP. Many other studies have also shown that MC is useful for detecting fundus diseases [68]. However, the limitation of MC is that there were no regular and continuous session visits arranged for the patients. Furthermore, it has been observed that the artifacts may affect the quality of images, thereby compromising disease detection.

G Cennamo et al. [69] used OCTA for the examination of the Choriocapillary Vascular Density (CNV) in CSR disease. The authors considered twelve OCTA eye scans from twelve patients for testing. Their results determined that the OCTA provided a better solution among all other techniques such as, FA, SD-OCT, and ICGA, because these techniques are not diagnostic. Therefore, the OCTA proved to detect CNV in CSR by a direct visualization and resulted in attaining 100% sensitivity and specificity. Furthermore, a number of recent researches have reported that OCTA has the best ability to detect the choriocapillaris vascular density [70]. Table 1 presents a summary of the aforementioned CSR detection techniques in an ascending order. The summaries of the intricate peer-reviewed articles comprise of algorithms used, number of datasets considered for the study and the effectiveness of the proposed results. Essentially, Table 1 provides a quick glance of CSR imaging technologies for the readers of this review.

**TABLE 1**
**SUMMARY OF CSR DETECTION BASED ON IMAGING TECHNOLOGIES**

| Reference # | Year | Authors | Dataset | Key Technique | Results |
|---|---|---|---|---|---|
| [43] | 2015 | M. M. Teussink et al. | 2 OCT eye scans images | OCTA | Moderately higher (median JI, 0.74) as well as moderate (median JI, 0.52) |
| [46] | 2015 | J. Y. Shin Velthoven et al. | 73 OCT eye scans | Optomap-AF | Visual acuteness (VA) 0.2 (range 0–1.3). |
| [48] | 2015 | E. Costanzo et al. | 33 OCTA eye scans of 32 consecutive patients. | OCTA with multimodal imaging | Mean BCVA was $0.23 \pm 0.25$ |
| [51] | 2016 | F. Rupesh Agrawal et al. | OCT scans of 78 eyes of 39 patients | Image binarization | Mean best-corrected optical acuteness $0.31 \pm 0.38$ |
| [54] | 2016 | S. Weng Xiang et al. | 70 OCTA eye scans of patients | Optical-coherence tomographic angiography (OCTA) | 97.77%, 100%, 93.33% Accuracy, sensitivity and specificity respectively. |
| [56] | 2016 | Michael R et al. | 16 patients OCT images | slit-lamp bio microscopy and fundus photography | Efficiency |
| [59] | 2016 | Z. W. Jelena Novosel et al. | 74 retinal OCT pictures | max-flow optimization algorithm | Dice coefficient, TPR and FPR 0.96,95% and 1%, respectively |
| [61] | 2017 | M.Wu Xiang et al. | 37 retinal SD-OCT | | 95 % Accuracy. |
| [63] | 2019 | H. J. Chien et al. | 20 OCT scan images | Balloon Snake Algorithm | Performance: 30 min faster than manual segmentation |
| [37] | 2019 | K. Gao, et al. | 200 B-scan images | Graph theory and B-scan | Suggests absolute thickness differences parallel to manual segmentation $3.68\pm2.96$ μm and $5.84\pm4.78$ μm |
| [66] | 2019 | SG Odaibo et al. | 90 time-domain OCT images | Cloud-based-mobile Artificial Intelligence (AI) platform | 82.5% to 97% sensitivity and specificity from 52% to 100% |
| [67] | 2020 | M. LU HE et al. | 75 SD-OCT eyes scans of 69 patients including disease eyes | Multicolored imaging (MC) and Color Fundus Photography (CFP) | 92.0% as a detection rate. |
| [69] | 2020 | G Cennamo et al. | OCTA images scans of 12 eyes | Optical-Coherence Tomography Angiography | CVD of controls resulted in significant higher in Group 2 at baseline (whole, parafovea and fovea p<0.05). |

## 10) ARTIFICIAL INTELLIGENCE BASED DETECTION OF CSR

Artificial Intelligence (AI) techniques are divided into three different phases: training, validation and testing for arrangement of images [71]. The most advanced AI techniques include Machine Learning (ML) and its more recent variant known as Deep Learning (DL). Machine Learning (ML) is further categorized into supervised, unsupervised and reinforcement learning. In case of supervised learning, the targeted variables are predicted from a prearranged indicator (known as the Objective Function) for simplicity. Taking advantage of arrangement of factors, generation of the Objective Function guides contribution to anticipated targets. The dataset training phase keeps iterating until the model attains an ideal degree of precision. The examples of supervised learning techniques are Regression & Logistic Regression, KNN, Decision Tree, and Random Forest, etc. In unsupervised learning, the prediction of the objective or desired variable is carried out. It is employed for the collecting population in various groups, which is broadly used for fragmenting purposes. The examples of unsupervised learning include K-means and Apriori algorithm. In reinforcement learning, the machine is normally trained to sort and categorize explicit choices. It is programmed to strategically learn itself when introduced to a certain circumstance and up-skill reliably with trial and error experimentation. This allows the algorithm to experience and attempt to catch the ideal data to attain the precise business choices. An example of reinforcement learning includes Markov Decision Process. Figure 9 shows the methodology of Machine Learning (ML) model in which fundus retinal image is used for classification of CSR.

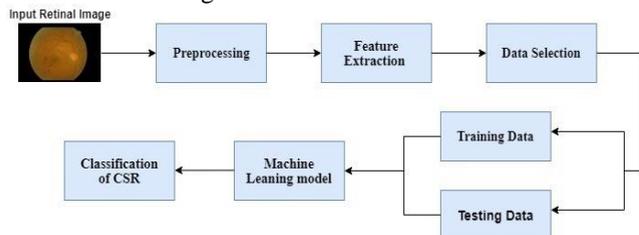

**FIGURE 9:** ML model for classification of CSR

The process starts with inputting a retinal image, which is preprocessed for any anomaly detection or better visualization for the next stages. After preprocessing, the next stage is feature extraction in which various parts and features of the eye scan are extracted. Then specific parts of the scan are selected for training or testing purposes. If the image is inputted specifically for the training purpose, it is processed through various algorithmic stages to boost up the ML model, thereby obtaining a particular pre-classified CSR related image. Contrarily, if the image was initially inputted for testing purposes, then it goes through the predefined algorithmic stage to test the model for its validation. In this way, the model is first trained using a large dataset of images, and then tested using more data. Once the algorithmic model goes through hundreds of thousands of images, it is trained to determine any CSR diseases in future inputted images.

Deep Learning (DL) is an advanced version of ML based on neural networks capable of unsupervised learning from an unstructured and unlabeled dataset. In recent studies, DL algorithms are vastly used by researchers and scientists in almost every field including image recognition and automated disease detection. In ophthalmology, DL has been applied to fundus images, optical intelligibility tomography and visual fields, accomplishing robust classification execution in the discovery of CSR [72]. Figure 10 shows the methodology of deep learning model in which fundus retinal image is used for classification of CSR.

## 11) CONVOLUTIONAL NEURAL NETWORK (CNN)

The Convolutional Neural Network (CNN) is a deep learning (DL) architecture that requires unstructured data (such as scanned images) as inputs, that allocate learning weights and biases to different viewpoints/entities in the image and have the option to separate one from the other based on their significance [73].

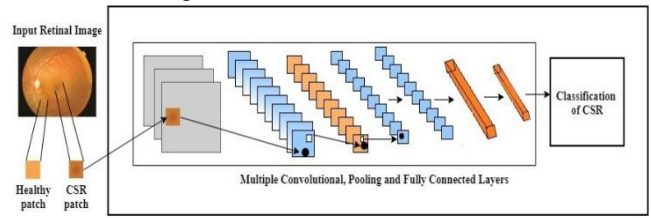

**FIGURE 10:** CNN architecture for classification of CSR

The development and advancement of the CNN declares the capacity to learn automatically specific to a training dataset as per the predictive modeling problem. The image classification can be the best example of CNN and their outcomes can easily be determined to specific features anywhere on input images [74]. Figure 10 depicts the methodology of CNN architecture in which fundus retinal image is used for classification of CSR. It is evident that the input eye scan image goes through multiple convolutional layers which have the self-learning ability to focus on the CSR patch. After the final CNN layer, the model can accurately segment the image and determine the exact location and extent of the CSR patch. Similar to this patch, numerous other features related to various anomalies are focused in different DL models. In the following paragraphs, the use of various ML/DL algorithms in automatic detection of CSR disease have been reviewed, along with brief description of the results and shortcomings.

Hassan B et al. [75] had proposed the automated and rational algorithm to detect CSR through OCT images. In this methodology, the input OCT scan has been preprocessed by de-noising and enhancing the image and then ILM and choroid are segmented by applying optimal thresholding. Feature extraction has been performed and the training dataset is passed through SVM classifier and the required classification is achieved. The algorithm detects images by applying a Support Vector Machine (SVM). In this mechanism, five different characteristics were extracted from a dataset gathered from different hospitals. The performance of the SVM algorithms has been observed better than others as proposed in [76]. The authors algorithm truly distinguished 88 out of 90 subjects attaining 97.77% accuracy, 100% sensitivity and 93.33% specificity [75].

A. M. Syed et al. [77] proposed the rational technique to overcome the CSR disease problem. The methodology consists of three phases. In the first phase, the robust reconstruction of 3-D OCT retinal surfaces was planned. In the second phase 2 feature sets are formulated, one is thickness profile and the other is cyst fluid. The retinal subject is classified using SVM classifier. The lodged algorithm worked on multiple OCT images for the detection, whereas several researchers worked on OCT and fundus images [78], [79]. The sample size of 18 OCT images was used as a dataset. After conducting the experiments and calculations, authors achieved the accuracy of 98.88%, sensitivity of 100%, and specificity of 96.66%. Furthermore, the authors reported that their proposed algorithm is faster and more robust in detecting CSR compared to the state-of-the-art researches. Finally, the authors claimed to utilize the proposed algorithm for diagnosing other macular disorders such as, AMD and glaucoma.

S. Khalid et al. [80] proposed a decision support system that instinctively aimed to dredge up CSR from retinal images using the SVM classifier. The methodology works by preprocessing i.e. sparse de-noise the input image and then segmentation is applied. This segmentation results into retinal layers whose retinal thickness profile is generated. Features extracted are passed to the SVM classifier.

The dataset contained 90 OCT scans of 73 patients for experimental purpose. The system ratings of 99.92%, 100%, and 99.86% attained accuracy, sensitivity, and specificity respectively [80]. In addition, the authors stated that any other intricate research works had utilized their OCT pictures to distinguish retinal variations from the norm [81], [82].

D. Xiang et al. [83] proposed the random forecast classifiers incorporating a live wire algorithm. This algorithm intends to find surfaces in optical coherence tomography. The sample SD-OCT images of 24 spectral domains suffering with CSR from various hospitals were used to conduct experiments. The authors used both manual and automated methods to detect the CSR. The authors claimed that the results obtained from their algorithm outperformed the contemporary methods. Even though, the proposed strategy has been utilized in OCT imaging with CSR, the underlying algorithm could be efficiently utilized in different origins of OCT pictures. B. Hassan et al. [84] proposed the CAD-based fully automated algorithm using the KNN classifier to detect CSR. The authors utilized a dataset of 80 retinal subjects to conduct the experimental study. The proposed algorithm was assumed to extract retinal layers through 3D features. The proposed algorithm supports practitioners and doctors while detecting CSR efficiently and accurately. The results had achieved 100% efficiency in diagnosing the CSR. The numerous strategies eventually resulted in the acquisition of eye checks as it may be like OCT imaging and has emerged as one of the best modern methods [85], [86]. Despite its good features, the algorithm can be modified more to attain the exact images with precision. In addition, this methodology can equally be reached out to other retinal pathologies in the future.

Most of the CSR detection methods commonly impart the process of segmentation. However, the authors Z Ji et al. [87] proposed an algorithm that detects CSR without segmentation of retinal layers named as '*Hessian-based Aggregate comprehensive Laplacian of Gaussian algorithm*'. The ensuing framework is now commonly used for the automatic segmentation of Neurosensory Retinal Detachments (NRD). The 23 samples of longitudinal SD-OCT scans were collected for the testing phase. The execution and implementation process require filtration of B-scans into little blob areas dependent on local convexity by amassing the log-scale-standardized convolution responses of every individual gLoG filter. The feature vectors were sustained in the direction of unsupervised clustering algorithm. The testing results showed excellent performance as compared to the other manual methods. Authors achieved 95.15%, 93.65% and 94.35% as a mean true positive volume fraction, positive predictive and dice similarity coefficient respectively. Various state-of-the-art models were proposed in [88]–[91] for detecting sub-retinal fluids in CSR, most of which are retinal layer segmentation dependent. Therefore, the introduced methodology can deliver precise division outcomes that are exceptionally reliable with the argument without retinal layer separation. This technique may provide dependable NRD segmentations to SD-OCT images and can be helpful for clinical analysis. The future direction of the proposed strategy essentially centers on the expansion of 3-D segmentation.

DL plays a vital role in detecting sub-retinal fluids from SD-OCT images and improving its boundaries day by day. J De Fauw et al. [92] had proposed a self-activating and self-learning technique to segment retinal fluid from SD-OCT CSR using a DL algorithm known as '*deep segmentation network'*. This network uses a 3-D U-Net architecture to interpret abraded OCT into tissue map. This algorithm was trained on a very large dataset of 14,884 tissue-maps with committed diagnosis and was later tested on 887 SD-OCT scanned images. On the other hand, several researchers have worked on a single end-to-end black-box network that needs indefinite number of OCT scans [10]. This algorithm uses a group of diverse segmentation and classification models: (i) CNV (ii) MRO (iii) Normal (iv) CSR (V) Mac hole. This also incorporates four DL phases: the digital OCT scans, segmentation network, tissue segmentation map, and classification network. The 'black-box' issue is distinguished as a hindrance to the use of DL within healthcare systems. The system structure intently coordinates the clinical dynamic procedure, distinguishing decisions independently from the subsequent referrals based on the scan. This practice permits a clinician to investigate and envision an interpretable division instead of merely giving a conclusion and referral recommendation. The proposed algorithm has achieved a total error rate (1- accuracy) of 5.5% AUC on the referral decisions. In [93], K. Gao et al. proposed the innovative image-to-image, double-branched, fully-in-convolutional networks (DA-FCN) for segmentation of retina fluid from SD-OCT images. Many researchers worked on the same disease but with other methods like Montuoro et al., proposed a 3D graph search technique for the detection of CSR [94]. First the dataset is extended by applying mirroring technique and then shallow coarse is learnt by designing double branched structures. In the end, area loss is joint with softmax loss in order to learn features that are more useful.

The proposed technique was prospective to provide a measurable evaluation of NRD just as PED. The sample of 52 SD-OCT test images dataset was acquired from 35 eye scans of 35 patients. The results depicted that 94.3% of true positive value fraction, 94.3% dice similarity efficient, and 96.4% positive predicative value was attained. Moreover, the proposed method had excessive achievements in TPVF and DSC compared to the other related methods given in the literature. These attainments can be enhanced and improved to develop features that are more distinguished by introducing loss factors, which can segment retinal fluid efficiently and precisely.

In addition to the aforementioned causes of CSR, Maculopathy plays a vital role in the blindness of the human eye. It has been diagnosed by automated systems like OCT imaging, and it may act as harmful object for the human eye. B. Hassan et al. [95] had proposed to develop a rational technique for segmentation and grading according to medical criteria. An algorithm called Support Vector Machine (SVM) was used, which works in six different phases: (i) data acquisition (ii) preprocessing (iii) retinal layer segmentation (iv) retinal fluid detection (v) feature set formulation, and (vi) classification. The clinical review of CSR and ME alongside the seriousness examination and 3D profiling utilizing OCT images were also presented.

The dataset of 30 OCT test images was collected from seventy-three different patients. Several researchers have worked on retinal fluid segmentation like computer aided diagnosis method by B. Hassan et al. [95] for self-activating identification of idiopathic CSR. The proposed framework first performs preprocessing one the image, then retinal layer segmentation is performed using Tensor Graph and then retinal fluid is detected. Features extracted are passed to SVM classifier and disease grading is performed. The precision of the proposed work was reported as the true affirmative rate as well as the true negative rate of 97.78%, 96.77%, and 100%, respectively. The technology of OCT imaging, when integrated with ML algorithms, turns out to be advantageous. R V Teja et al. [96] had proposed the integrated random forest classifier and DeepLab algorithm for the detection of CSR. The sample sizes of 768 B-scan tests were collected. The phases of classification and segmentation were determined through a number of famous frameworks known as the XGBoost classifier, the random forest classifier and the DeepLab. The testing score on OCT images was achieved as the mean dice of 86.23% as compared to the manual delineations made by the trained experts. Recent studies [97] also show that DeepLab set another cutting edge on a several benchmark semantic image segmentation datasets. Y. Zhen et al. [98] proposed the mechanized methodology based on DL architecture termed as *InceptionV3* to recognize CSR portrayed on color fundus images. The sample size for training and testing was based on 2504 OCT-images that were pre-processed and normalized. The results and experiments concluded that the proposed algorithm is dependent on DL may evaluate CSR portrayed on color fundus in a moderately solid manner. The authors claimed that their results showed that the AUC was 0.927 (95% CI: 0.895-0.959) and the p-value was less than 0.001. Moreover, the cutoff threshold was 0.5 and the accuracy was 85.7%. The restricted dataset was used for creating model, but the trials showed the special quality of profound learning innovation. Future translational exertion is alluring to make such a tool accessible to the clinical practice for beginning CSR screening results in early stage diagnosis. Several researchers have worked on color fundus images but these images were observed to be related to diabetic retinopathy and diabetic macula edema [10] but very limited work have been conducted on CSR. The combination of DL algorithm and fundus photography can be very efficient for the ophthalmologists in their clinical diagnosis, which may reduce unnecessary usage of OCT-images and fluorescein photography. Finally, it must be pointed out that this study contains some limitations which include the less variety of images and the effects of image artifacts on CSC assessment Y. Ruan et al. [99] proposed the Fully Convolutional Networks (FCN) with multiphase level set named (FCN-MLS) for segmenting the boundaries in retinal images. The system proposed by the authors has been partitioned into three main phases: (i) Pre-processing (ii) FCN for layer boundaries segmentation (iii) Distance regularization level set (DRLS). The sample size of 10 SD-OCT test datasets was collected, thereby making it 120 B-scan images. The experimental results obtained a gross average difference of absolute boundary location and the gross average difference of boundary thickness, which were $5.88 \pm 2.38\mu m$ and $5.81 \pm 2.19\mu m$, respectively. Various researchers have worked on retinal layer segmentation using Artificial Intelligence (AI) and it worked as the state-of-art methodology [100]. TJN Rao et al. [101] proposed the segmentation algorithm based on neural networks to detect CSR. The capacity of a convolutional neural system to highlight the separation of unpretentious spatial variety that converts retinal fluid into division task [102]. This method will find and segment the retinal fluid from OCT images and reduce the impact of noise at background. It is based on two phases: (i) pre-processing phase and (ii) fluid segmentation phase. Dataset for experiments consists of 15 OCT images obtained from CSR patients. The experimental results with dice rate of 91%, precision 93%, and recall 89% were reported. Furthermore, the authors claimed that the development of hybridization of neural network and feature extraction techniques would be used as a future work for better segmentation of CSR.

S.A.E. Hassan et al. [6] proposed a method for fully automated Central Serous Retinopathy (CSR) detection using deep CNN. In the developed framework, the pre-processing phase enhanced the image quality and eliminated noise. The three pre-trained classifiers (AlexNet, ResNet-18, and GoogleNet) have been used for classification. The outcomes indicate that AlexNet achieved the highest accuracy 99.64%, precision 98.91%, recall 100%, and f1-score 99.45% on publicly available OCT images dataset.

Pawan et al. [103] proposed an enhanced SegCaps architecture based on Capsule Networks for the segmentation of SRF using CSCR OCT images. The developed method was outperformed in UNet architecture and reduced parameters to be trained by 54.21%. Furthermore, it was also reduced the computation complexity of SegCaps by 37.85%, with comparison to other method's performance.

Chen et al. [104] presented a study consisting of 2104 FFA images from FFA sequences of 291 eyes, including 137 right eyes and 154 left eyes from 262 patients. The attention gated Network (AGN) has been used for the segmentation of leakage points and UNet was used for optic disk (OD) and macula region segmentation. The results showed that the detection was perfectly matched 60.7% in the test set. After utilizing eliminated method, the accurate detection cases increased 93.4% and the dice on lesion level was 0.949, respectively. Future work includes the improvement of Computational Complexity.

Table 2 presents a list of the research papers considered for this review paper that particularly uses Machine/Deep learning technologies for CSR detection. This table comprises scholar name, year of publication, proposed algorithm to solve the problem, suggested methodology for the problem, dataset used, and outcomes respective to sensitiveness, specificity, accuracy, and region fall under ROC curve.

**TABLE 2**
**SUMMARY OF CSR DETECTION BASED ON MACHINE/DEEP LEARNING IN CSR DETECTION**

| Reference # | Year | Authors | Dataset | Key Technique | Results |
|---|---|---|---|---|---|
| [75] | 2016 | Hassan B et al. | 34 OCT scans | SVM classifier | 97.77%, 100%, 93.33% accuracy, sensitivity and specificity respectively |
| [77] | 2016 | A. M. Syed et al | 90 OCT volumes | SVM classifier | 98.88%, 100%, and 96.66% accuracy, sensitivity and specificity respectively |
| [80] | 2017 | S. Khalid et al. | 90 OCT scans of 73 patients | Multi layered SVM classifier | 99.92%, 100%, and 99.86% Accuracy, sensitivity and specificity respectively. |
| [83] | 2018 | D. Xiang et al. | 6 macula-centered SD-OCT images | Random Forest and Live wire algorithm | Sum of mean differences of absolute boundary and thickness positioning differentiated to manual segmentation results: $3.68 \pm 2.96$ μm and $5.84 \pm 4.78$ μm. |
| [84] | 2018 | B. Hassan et al. | 80 CSR retinal subjects | CAD-based fully automated algorithm using the KNN classifier | Value of k = 5, achieving 99.8 % accuracy 100 % Efficiency |
| [87] | 2018 | Z Ji et al. | 23 longitudinal SD-OCT scans. | Hessian-based Aggregate comprehensive Laplacian of Gaussian algorithm | Sum of true positive volume fraction (95.15%), positive predicative value (93.65%) and dice similarity coefficient (94.35%). |
| [92] | 2018 | J De Fauw et al. | 14,884 training tissue maps and 887 SD-OCT scans. | DL algorithm known as 'deep segmentation network | Mean error rate (1- accuracy) 5.5% AUC on referrals decision. |
| [93] | 2019 | K. Gao et al. | 52 SD-OCT images dataset of 35-eyes of 35 subjects | DA-FCN (fully convolutional networks) | True positive value fraction 94.3%, dice similarity efficient 94.3%, and positive predicative value 96.4%. |
| [95] | 2019 | B. Hassan et al. | 30 OCT images of seventy-three patients | SVM classifier | Accuracy, true positive rate and true negative rate of 97.78%, 96.77%, 100% separately |
| [96] | 2019 | R V Teja et al. | OCT images | Random forest classifier and DeepLab algorithm | Average Dice Sore of 86.23% |
| [98] | 2019 | Y. Zhen et al. | 2504 Fundus images | DL architecture termed as InceptionV3 | AUC 0.927, p-value < 0.001 cutoff threshold 0.5, accuracy 85.7% |
| [99] | 2019 | Y. Ruan et al | 10 SD-OCT subjects | FCN-MLS | Gross average difference of absolute boundary location and thickness were ($5.88 \pm 2.38$μm) and ($5.81 \pm 2.19$μm) respectively. |
| [101] | 2019 | TJN Rao et al. | 15 OCT volumes got from CSR patients | Segmentation algorithm based on neural networks to detect CSR | Dice rate 91%, precision 93%, and recall 89% |

| [6] | 2021 | S. A. E. Hassan et al. | 309 OCT images | AlexNet, ResNet and GoogleNet | 99.64% classification accuracy of AlexNet |
| [103] | 2021 | Pawan, S. J., et al. | OCT images | SegCaps (Capsule Network) | Reduction of trainable parameters by 54.21% and reduction of computational complexity by 37.85%. |
| [104] | 2021 | Chen, Menglu, et al. | 291 FFA images | U-Net and Attention Gated Network (AGN) | Accurate detection cases increased to 93.4% |

## III. DISCUSSION

After a thorough study of 29 relevant articles, section II looked at several CSR disease detection methodologies and algorithms. There were two broad categories for this approach: a) Detection of CSR based on classical imaging technologies, and b) Detection of CSR based on Machine/Deep Learning techniques. Table 1 & 2 highlight the key information about the relevant publications regarding both of these categories. This information includes the authors' list, the size of datasets, the underlying algorithms, and results summary. The Artificial Intelligence (AI) techniques mentioned in this review are primarily based on ML & DL approaches. The ML approach is further classified into three sub-categories, namely, supervised learning (Regression, Decision Tree, Random Forest, KNN, and Logistic Regression etc.), the unsupervised learning (Apriori algorithm, K-means) and the Reinforcement Learning (Markov Decision Process). From a careful analysis, it can be determined that the utilization of the classical ML approach has yielded excellent results in accuracy, reliability, and speed for a given sample dataset to diagnose CSR.

On the other hand, the DL technique is quite beneficial for fundus image sorting procedures, and it yields better results accuracy results than the classical ML approach. However, the results largely depend on the size of datasets and the complexity of the underlying algorithms. This review provides a qualitative and quantitative assessment of the algorithms in both aforementioned categories (see Tables 1 & 2 for summary). Furthermore, this meta-analysis of previous researches has identified several algorithms such as OCTA algorithms and SVMs. The results are described in terms of accuracy, sensitivity, and error rates. From a careful analysis, it can be determined that the KNN classifier (in the category of ML) has proposed to attain the best accuracy, followed by the SVM classifier has attained high accuracy and double-branched algorithm as well as area Constraint Fully Convolutional Networks (DA-FCN) with moderate accuracy. In case of DL, the Mobile Artificial Intelligence Platform reported to have attained maximum accuracy, followed by Robust Reconstruction of 3-D OCT Retinal Surfaces, which attained 98.88% accuracy. Finally, the Split-Spectrum Amplitude-Decorrelation Angiography (SSADA) methods attained slightly moderate accuracy. The most significant measure to ascertain the reliability of results is the size of datasets in all cases. Some of the researchers claim that their results portray an accurate detection of the CSR disease based on their models trained under specific datasets. In most cases, the datasets are not large enough to properly train the models to accurately detect any future anomalies in scanned images fed to them for further analysis. Therefore, the reliability and size of datasets used to train the models must be major criteria for quantifying the accuracy of the results. Tables 1 & 2 give the size of datasets as well as the relevant accuracy of the results, and subsequently, a more accurately reported result based on smaller dataset should be considered as less reliable compared to a less accurate result based on a large dataset.

Another important parameter is to check the complexity and comprehensiveness of the algorithms used for detection. The comprehensiveness of algorithm means that it should take into account every type of anomalies for CSR detection. The DL algorithms have greater complexity due to their usage of Artificial Neural Networks (ANNs), which are composed of multiple layers through which the data is processed and transformed. Hence, they are more accurate at the cost of their complexity. The downside of these algorithms is that they are normally computationally intensive in nature, and they require a lot of processing time during the training and testing phases. Their cost of reliability and accuracy is computational time, especially if they are trained and tested over a huge dataset.

Another major limitation in most of these ML/DL models is that they have been trained & tested on proprietary datasets that are not publicly available. This raises an important question of their integrity and authenticity. There are very limited publicly available datasets; in fact, there are only two such datasets. Most of the models are trained on private datasets, which are proprietary to their owner institutions. Therefore, more public datasets should be available for the research community to train & test their AI models. This practice will not only encourage the researchers to access the datasets easily, but it will also help to standardize the models for the CSR disease detection. In this manner, the research community can develop a generic standard detection model.

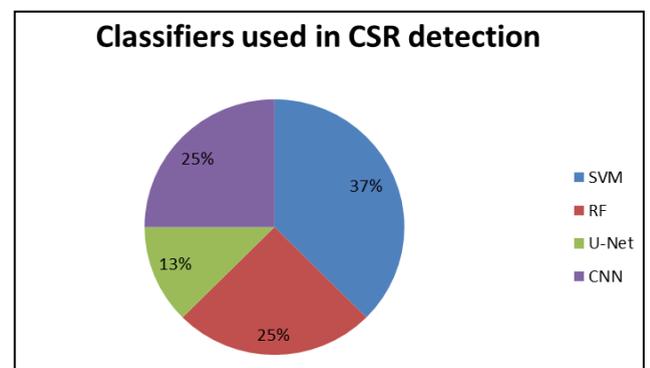

**FIGURE 11:** Classifiers used in the literature for CSR detection

Figure 12 depicts the classification accuracy attained by ML/DL classifiers and figure 13 shows total no of images used in the literature with respect to each classifier.

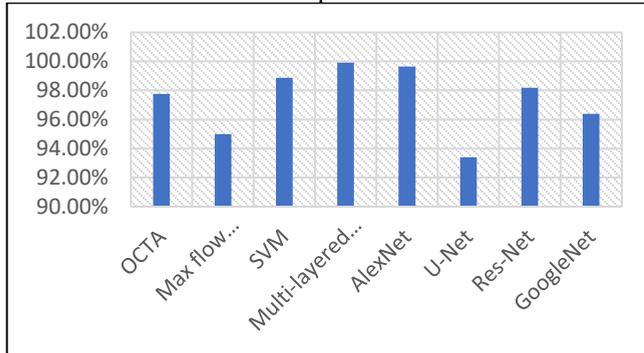

**FIGURE 12:** Comparison of literature based on classification accuracy

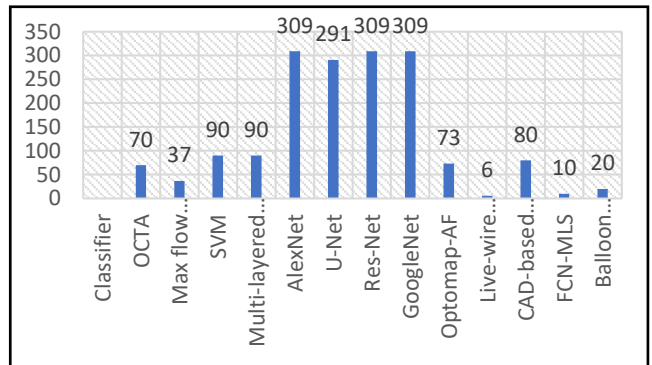

**FIGURE 13:** Comparison of literature based on validation data images

## IV. CONCLUSION

Traditionally, the CSR disease due to retinal fluids was detected by experienced ophthalmologists after manually examining the scanned images of retina. However, this approach is time-consuming, inaccurate and unreliable. With the recent advancements in technology, the CSR has been automatically detected with the aid of various AI-based models. Although the optimal detection of CSR is still a big challenge for the researchers, these intelligent techniques have become more popular and effective compared to the manual detection approaches. The recent literature found that numerous intelligent algorithms on OCT images are utilized for the CSR detection, and the experts of segmentation procedures carefully control this process. However, it is still subjected to human error. Therefore, the latest technology relies on AI based algorithms, which include ML & DL. These algorithms form accurate detection models, which are trained & tested on the legacy datasets. After proper testing and tuning, the researchers and physicians for the automatic CSR detection and diagnosis utilize these models. In recent times, these models are becoming more and more accurate and reliable. Furthermore, there are numerous commercially available products using AI based models. Moreover, the performance of these models must be precisely analyzed with the standard measurements and benchmarks.

In this study, a detailed review of automatic detection of CSR using AI techniques has been presented. These automatic detection methods assist in detecting the disease of vision loss and blindness as a result of CSR. These methods were categorized into classical imaging and ML/DL based techniques. Furthermore, this study encompassed the retinal datasets, strategies proposed for CSR assessment and accuracy. The additional retinal neovascularization, hemorrhages, miniaturized scale aneurysm, and exudates were also been highlighted for dataset of CSR detection. The roles of assessment measurements for Computer Aided Diagnosis (CAD) frameworks were deeply examined. The significance of each ML and DL methodology was ascertained to perform a better analysis of CSR for research community and ophthalmologists. A number of recent research articles were studied and examined to compare their respective algorithms, usage of their datasets and the accuracy of their results. A detailed analysis showed that the advanced ML/DL models are being employed for accurate, reliable, and rapid detection of CSR. The algorithm introduced by Hassan et al [84] is the most promising algorithm as it gives 99.8% accuracy. This open source CSR detection model will provide agility, sustainability flexibility and cost effectiveness, thus it can serve the mankind in a much better way.

Detection of CSR is evolving, using Machine Learning/Deep Learning Algorithms and imaging technologies. However, it is still mattered of controversy, and to date studies that have been published so far depends on the old methodologies and are unrestrained. Failure to detect and diagnose CSR may result in complete vision loss. Although, researches showing promising results but more research is required on publically available dataset and improving the computational complexity. Therefore, further studies are warranted to address this issue, improving the patient's care.

This paper provides a comprehensive review than existing review articles on existing methodologies to detect CSR through ML/DL techniques. Additionally, the limitations of these methodologies as well as the key suggestions have been mentioned to cope up with the limitations for the assistance of research community. Moreover, an open question for future research can be to create an open-source CSR detection model based on Deep Learning technologies, which is accessible to the research community to test it on their proprietary datasets or publically available ones.

### CONFLICT OF INTEREST
The authors of this review paper have no conflict of interest.

### ACKNOWLEDGEMENT
This work was supported by the Artificial Intelligence & Data Analytics Lab (AIDA), Prince Sultan University, Riyadh, Saudi Arabia in collaboration with Riphah Artificial Intelligence Research (RAIR) Lab, Riphah International University, Faisalabad Campus, Pakistan. Authors also appreciate Prince Sultan University support for paying APC for this publication.

## REFERENCES

[1] C. E. Willoughby, D. Ponzin, S. Ferrari, A. Lobo, K. Landau, and Y. Omidi, "Anatomy and physiology of the human eye: Effects of mucopolysaccharidoses disease on structure and function - a review," *Clinical and Experimental Ophthalmology*. 2010, doi: 10.1111/j.1442-9071.2010.02363.x.

[2] A. A. Dahl and T. R. Gest, "Retina Anatomy," *Medscape Reference*. pp. 1–4, 2015, doi: "https://www.nvisioncenters.com/education/eye-structure-anatomy/.

[3] M. U. Akram, S. Akbar, T. Hassan, S. G. Khawaja, U. Yasin, and I. Basit, "Data on fundus images for vessels segmentation, detection of hypertensive retinopathy, diabetic retinopathy and papilledema," *Data in Brief*, 2020, doi: 10.1016/j.dib.2020.105282.

[4] S. Akbar, T. Hassan, M. U. Akram, U. U. Yasin, and I. Basit, "AVRDB: Annotated dataset for vessel segmentation and calculation of arteriovenous ratio," in *Proceedings of the 2017 International Conference on Image Processing, Computer Vision, and Pattern Recognition, IPCV 2017*, 2017.

[5] S. Akbar, M. Sharif, M. U. Akram, T. Saba, T. Mahmood, and M. Kolivand, "Automated techniques for blood vessels segmentation through fundus retinal images: A review," *Microscopy Research and Technique*. 2019, doi: 10.1002/jemt.23172.

[6] S. A. E. Hassan, S. Akbar, S. Gull, A. Rehman, and H. Alaska, "Deep Learning-Based Automatic Detection of Central Serous Retinopathy using Optical Coherence Tomographic Images," in *2021 1st International Conference on Artificial Intelligence and Data Analytics (CAIDA)*, 2021, pp. 206–211, doi: 10.1109/CAIDA51941.2021.9425161.

[7] T. A. Soomro *et al.*, "Deep Learning Models for Retinal Blood Vessels Segmentation: A Review," *IEEE Access*. 2019, doi: 10.1109/ACCESS.2019.2920616.

[8] M. D. Abràmoff *et al.*, "Improved automated detection of diabetic retinopathy on a publicly available dataset through integration of deep learning," *Investigative Ophthalmology and Visual Science*, vol. 57, no. 13, pp. 5200–5206, Oct. 2016, doi: 10.1167/iovs.16-19964.

[9] A. A. van der Heijden, M. D. Abramoff, F. Verbraak, M. V van Hecke, A. Liem, and G. Nijpels, "Validation of automated screening for referable diabetic retinopathy with the IDx-DR device in the Hoorn Diabetes Care System," *Acta Ophthalmologica*, vol. 96, no. 1, pp. 63–68, Feb. 2018, doi: 10.1111/aos.13613.

[10] V. Gulshan *et al.*, "Development and validation of a deep learning algorithm for detection of diabetic retinopathy in retinal fundus photographs," *JAMA - Journal of the American Medical Association*, 2016, doi: 10.1001/jama.2016.17216.

[11] A. Tufail *et al.*, "An observational study to assess if automated diabetic retinopathy image assessment software can replace one or more steps of manual imaging grading and to determine their cost-effectiveness," *Health Technology Assessment*, vol. 20, no. 92, pp. 1–72, Dec. 2016, doi: 10.3310/hta20920.

[12] R. Rajalakshmi, R. Subashini, R. M. Anjana, and V. Mohan, "Automated diabetic retinopathy detection in smartphone-based fundus photography using artificial intelligence," *Eye*, vol. 32, no. 6, pp. 1138–1144, Jun. 2018, doi: 10.1038/s41433-018-0064-9.

[13] M. Bhaskaranand *et al.*, "Automated Diabetic Retinopathy Screening and Monitoring Using Retinal Fundus Image Analysis," *Journal of Diabetes Science and Technology*, vol. 10, no. 2, pp. 254–261, Mar. 2016, doi: 10.1177/1932296816628546.

[14] T. A. Soomro, J. Gao, T. Khan, A. F. M. Hani, M. A. U. Khan, and M. Paul, "Computerised approaches for the detection of diabetic retinopathy using retinal fundus images: a survey," *Pattern Analysis and Applications*. 2017, doi: 10.1007/s10044-017-0630-y.

[15] S. Akbar, M. U. Akram, M. Sharif, A. Tariq, and U. ullah Yasin, "Arteriovenous ratio and papilledema based hybrid decision support system for detection and grading of hypertensive retinopathy," *Computer Methods and Programs in Biomedicine*, vol. 154, pp. 123–141, Feb. 2018, doi: 10.1016/j.cmpb.2017.11.014.

[16] S. Akbar, M. U. Akram, M. Sharif, A. Tariq, and S. A. Khan, "Decision support system for detection of hypertensive retinopathy using arteriovenous ratio," *Artificial Intelligence in Medicine*, vol. 90, pp. 15–24, Aug. 2018, doi: 10.1016/j.artmed.2018.06.004.

[17] S. Akbar, M. U. Akram, M. Sharif, A. Tariq, and U. ullah Yasin, "Decision Support System for Detection of Papilledema through Fundus Retinal Images," *Journal of Medical Systems*, vol. 41, no. 4, p. 66, Apr. 2017, doi: 10.1007/s10916-017-0712-9.

[18] D. Hua *et al.*, "Retinal Microvascular Changes in Hypertensive Patients with Different Levels of Blood Pressure Control and without Hypertensive Retinopathy," *Current Eye Research*, pp. 1–8, Jun. 2020, doi: 10.1080/02713683.2020.1775260.

[19] H. Bogunovic *et al.*, "Prediction of Anti-VEGF Treatment Requirements in Neovascular AMD Using a Machine Learning Approach," *Investigative*

*Opthalmology & Visual Science*, vol. 58, no. 7, p. 3240, Jun. 2017, doi: 10.1167/iovs.16-21053.

[20] M. Rohm *et al.*, "Predicting Visual Acuity by Using Machine Learning in Patients Treated for Neovascular Age-Related Macular Degeneration," *Ophthalmology*, vol. 125, no. 7, pp. 1028–1036, Jul. 2018, doi: 10.1016/j.ophtha.2017.12.034.

[21] M. Treder, J. L. Lauermann, and N. Eter, "Automated detection of exudative age-related macular degeneration in spectral domain optical coherence tomography using deep learning," *Graefe's Archive for Clinical and Experimental Ophthalmology*, vol. 256, no. 2, pp. 259–265, 2018, doi: 10.1007/s00417-017-3850-3.

[22] C. González-Gonzalo *et al.*, "Evaluation of a deep learning system for the joint automated detection of diabetic retinopathy and age-related macular degeneration," *Acta Ophthalmologica*, vol. 98, no. 4, pp. 368–377, Jun. 2020, doi: 10.1111/aos.14306.

[23] E. Vaghefi, S. Hill, H. M. Kersten, and D. Squirrell, "Multimodal Retinal Image Analysis via Deep Learning for the Diagnosis of Intermediate Dry Age-Related Macular Degeneration: A Feasibility Study," *Journal of Ophthalmology*, vol. 2020, pp. 1–7, Jan. 2020, doi: 10.1155/2020/7493419.

[24] T. R. V. Bisneto, A. O. de Carvalho Filho, and D. M. V. Magalhães, "Generative adversarial network and texture features applied to automatic glaucoma detection," *Applied Soft Computing*, vol. 90, p. 106165, May 2020, doi: 10.1016/j.asoc.2020.106165.

[25] D. C. Hood, Z. Z. Zemborain, E. Tsamis, and C. G. De Moraes, "Improving the Detection of Glaucoma and Its Progression: A Topographical Approach," *Journal of Glaucoma*, vol. 29, no. 8, pp. 613–621, Aug. 2020, doi: 10.1097/IJG.0000000000001553.

[26] P. Kaur and P. K. Khosla, "Artificial Intelligence Based Glaucoma Detection," Springer, Singapore, 2020, pp. 283–305.

[27] A. Sarhan, J. Rokne, and R. Alhajj, "Approaches for Early Detection of Glaucoma Using Retinal Images: A Performance Analysis," Springer, Cham, 2020, pp. 213–238.

[28] V. Mohammadzadeh *et al.*, "Macular imaging with optical coherence tomography in glaucoma," *Survey of Ophthalmology*, vol. 65, no. 6, pp. 597–638, 2020, doi: 10.1016/j.survophthal.2020.03.002.

[29] F. Li *et al.*, "Deep learning-based automated detection of glaucomatous optic neuropathy on color fundus photographs," *Graefe's Archive for Clinical and Experimental Ophthalmology*, vol. 258, no. 4, pp. 851–867, 2020, doi: 10.1007/s00417-020-04609-8.

[30] "Central Serous Retinopathy." https://www.slideshare.net/AnkushWeginwar/central-serous-retinopathy-73712617 (accessed Sep. 30, 2020).

[31] "Central serous retinopathy." https://www.medicalnewstoday.com/articles/320606#overview (accessed Sep. 30, 2020).

[32] M. B. Breukink *et al.*, "Chronic central serous chorioretinopathy: long-term follow-up and vision-related quality of life," *Clinical Ophthalmology*, vol. Volume 11, pp. 39–46, Dec. 2016, doi: 10.2147/OPTH.S115685.

[33] "Central Serous Retinopathy overview," 2013. [Online]. Available: https://www.rcophth.ac.uk/wp-content/uploads/2014/08/Focus-Winter-2013.pdf.

[34] G. Manayath, R. Ranjan, S. Karandikar, V. Shah, V. Saravanan, and V. Narendran, "Central serous chorioretinopathy: Current update on management," *Oman Journal of Ophthalmology*, vol. 11, no. 3, p. 200, Sep. 2018, doi: 10.4103/ojo.OJO_29_2018.

[35] M. A. Abouammoh, "Advances in the treatment of central serous chorioretinopathy," *Saudi Journal of Ophthalmology*, 2015, doi: 10.1016/j.sjopt.2015.01.007.

[36] "Central Serous Retinopathy before and after Angiograms." .

[37] K. Gao, W. Kong, S. Niu, D. Li, and Y. Chen, "Automatic retinal layer segmentation in SD-OCT images with CSC guided by spatial characteristics," *Multimedia Tools and Applications*, vol. 79, no. 7–8, pp. 4417–4428, Feb. 2020, doi: 10.1007/s11042-019-7395-9.

[38] M. Cardoso, T. Arbel, A. Melbourne, and H. Bogunovic, "Fetal, Infant and Ophthalmic Medical Image Analysis: International Workshop, FIFI 2017, and 4th International Workshop, OMIA 2017, Held in Conjunction," 2017.

[39] "Normal Coloured Retinal OCT image." https://www.ceenta.com/patientresources/the-retina (accessed Oct. 01, 2020).

[40] "Central Serous Retinopathy Coloured Retinal OCT image."https://imagebank.asrs.org/Content/imageba%0Ank/ped-WITH-SRF.jpg/imagefull;max$643,0.ImageHandler (accessed Oct. 01, 2020).

[41] P. Gholami, P. Roy, M. K. Parthasarathy, and V. Lakshminarayanan, "OCTID: Optical coherence tomography image database," *Computers and Electrical Engineering*, vol. 81, 2020, doi: 10.1016/j.compeleceng.2019.106532.

[42] D. S. Kermany *et al.*, "Identifying Medical Diagnoses and Treatable Diseases by Image-Based Deep Learning," *Cell*, vol. 172, no. 5, pp. 1122-1131.e9, 2018, doi: 10.1016/j.cell.2018.02.010.

[43] M. M. Teussink *et al.*, "Oct angiography compared

to fluorescein and indocyanine green angiography in chronic central serous chorioretinopathy," *Investigative Ophthalmology and Visual Science*, 2015, doi: 10.1167/iovs.15-17140.

[44] N. Kitaya *et al.*, "Features of abnormal choroidal circulation in central serous chorioretinopathy," *British Journal of Ophthalmology*, 2003, doi: 10.1136/bjo.87.6.709.

[45] A. Scheider, J. E. Nasemann, and O. E. Lund, "Fluorescein and indocyanine green angiographies of central serous choroidopathy by scanning laser ophthalmoscopy," *American Journal of Ophthalmology*, 1993, doi: 10.1016/S0002-9394(14)73524-X.

[46] J. Y. Shin, H. J. Choi, J. Lee, and M. Choi, "Fundus autofluorescence findings in central serous chorioretinopathy using two different confocal scanning laser ophthalmoscopes : correlation with functional and structural status," 2015, doi: 10.1007/s00417-015-3244-3.

[47] N. E. Gross, R. F. Spaide, D. L. L. Costa, S. J. Huang, J. M. Klancnik, and A. Aizman, "INDOCYANINE GREEN ANGIOGRAPHY – GUIDED PHOTODYNAMIC THERAPY FOR TREATMENT OF CHRONIC CENTRAL SEROUS A Pilot Study," pp. 288–298, 2003.

[48] E. Costanzo *et al.*, "Optical Coherence Tomography Angiography in Central Serous Chorioretinopathy," *Journal of Ophthalmology*, 2015, doi: 10.1155/2015/134783.

[49] M. Q. Maftouhi, A. El Maftouhi, and C. M. Eandi, "Chronic Central Serous Chorioretinopathy Imaged by Optical Coherence Tomographic Angiography," *American Journal of Ophthalmology*, pp. 1–8, 2015, doi: 10.1016/j.ajo.2015.06.016.

[50] M. Antonio *et al.*, "Association of Choroidal Neovascularization and Central Serous Chorioretinopathy With Optical Coherence Tomography Angiography," vol. 02116, no. 8, pp. 899–906, 2015, doi: 10.1001/jamaophthalmol.2015.1320.

[51] R. Agrawal, J. Chhablani, K.-A. Tan, S. Shah, C. Sarvaiya, and A. Banker, "CHOROIDAL VASCULARITY INDEX IN CENTRAL SEROUS CHORIORETINOPATHY," *Retina*, vol. 36, no. 9, pp. 1646–1651, Sep. 2016, doi: 10.1097/IAE.0000000000001040.

[52] C. Brandl, H. Helbig, and M. A. Gamulescu, "Choroidal thickness measurements during central serous chorioretinopathy treatment," *International Ophthalmology*, 2014, doi: 10.1007/s10792-013-9774-y.

[53] L. Yang, J. B. Jonas, and W. Wei, "Choroidal vessel diameter in central serous chorioretinopathy," *Acta Ophthalmologica*, vol. 91, no. 5, pp. e358–e362, Aug. 2013, doi: 10.1111/aos.12059.

[54] S. Weng, L. Mao, S. Yu, Y. Gong, L. Cheng, and X. Chen, "Detection of choroidal neovascularization in central serous chorioretinopathy using optical coherence tomographic angiography," *Ophthalmologica*, vol. 236, no. 2, pp. 114–121, 2016.

[55] R. F. Spaide *et al.*, "Central Serous Chorioretinopathy in Younger and Older Adults," *Ophthalmology*, vol. 103, no. 12, pp. 2070–2080, Dec. 1996, doi: 10.1016/S0161-6420(96)30386-2.

[56] C. Wong *et al.*, "Optical Coherence Tomography of Central Serous Chorioretinopathy," *American Journal of Ophthalmology*, vol. 120, no. 1, pp. 65–74, 1995, doi: 10.1016/S0002-9394(14)73760-2.

[57] H. (Ed. . Min, *Stereo Atlas of Vitreoretinal Diseases*. Springer Nature, 2019.

[58] E. A. Swanson *et al.*, "In vivo retinal imaging by optical coherence tomography," *Optics Letters*, 1993, doi: 10.1364/ol.18.001864.

[59] J. Novosel, Z. Wang, H. De Jong, M. Van Velthoven, K. A. Vermeer, and L. J. Van Vliet, "LOCALLY-ADAPTIVE LOOSELY-COUPLED LEVEL SETS FOR RETINAL LAYER AND FLUID SEGMENTATION IN SUBJECTS WITH CENTRAL SEROUS RETINOPATHY Rotterdam Ophthalmic Institute , 1 Rotterdam Eye Hospital , Rotterdam , The Netherlands Quantitative Imaging Group , Departmen," pp. 702–705, 2016.

[60] X. Xu, K. Lee, L. Zhang, M. Sonka, and M. D. Abràmoff, "Stratified Sampling Voxel Classification for Segmentation of Intraretinal and Subretinal Fluid in Longitudinal Clinical OCT Data," *IEEE Transactions on Medical Imaging*, 2015, doi: 10.1109/TMI.2015.2408632.

[61] M. W. U. Englin *et al.*, "Three-dimensional continuous max flow optimization-based serous retinal detachment segmentation in SD-OCT for central serous chorioretinopathy," vol. 8, no. 9, pp. 4257–4274, 2017.

[62] D. C. Fernández, "Delineating fluid-filled region boundaries in optical coherence tomography images of the retina," *IEEE Transactions on Medical Imaging*, 2005, doi: 10.1109/TMI.2005.848655.

[63] H. Chien, C. Chang, H. Chien, and C. Chang, "Seminars in Ophthalmology Application of the Balloon Snake in the Volume Measurement of Subretinal Fluid in Central Serous Chorioretinopathy Application of the Balloon Snake in the Volume Measurement of Subretinal Fluid in Central Serous Chorioretinopathy," *Seminars in Ophthalmology*, vol. 0, no. 00, pp. 1–6, 2019, doi: 10.1080/08820538.2019.1640749.

[64] M. Wei, Y. Zhou, and M. Wan, "A fast snake model based on non-linear diffusion for medical image segmentation," *Computerized Medical Imaging and Graphics*, 2004, doi: 10.1016/j.compmedimag.2003.12.002.

[65] J. Ruiz-Medrano, M. Pellegrini, M. G. Cereda, M. Cigada, and G. Staurenghi, "Choroidal Characteristics of Acute and Chronic Central Serous Chorioretinopathy Using Enhanced Depth Imaging Optical Coherence Tomography," *European Journal of Ophthalmology*, vol. 27, no. 4, pp. 476–480, Jun. 2017, doi: 10.5301/ejo.5000796.

[66] S. G. Odaibo, M. MomPremier, R. Y. Hwang, S. Yousuf, S. Williams, and J. Grant, "Mobile artificial intelligence technology for detecting macula edema and subretinal fluid on OCT scans: Initial results from the DATUM alpha Study," *arXiv preprint arXiv:1902.02905*, 2019.

[67] L. He, C. Chen, Z. Yi, X. Wang, J. Liu, and H. Zheng, "CLINICAL APPLICATION OF MULTICOLOR IMAGING IN CENTRAL SEROUS CHORIORETINOPATHY," *Retina*, vol. 40, no. 4, pp. 743–749, Apr. 2020, doi: 10.1097/IAE.0000000000002441.

[68] L. Reznicek *et al.*, "Scanning laser 'en face' retinal imaging of epiretinal membranes," *Saudi Journal of Ophthalmology*, 2014, doi: 10.1016/j.sjopt.2014.03.009.

[69] G. Cennamo, C. Comune, F. Mirra, P. Napolitano, D. Montorio, and G. de Crecchio, "Choriocapillary vascular density in central serous chorioretinopathy complicated by choroidal neovascularization," *Photodiagnosis and Photodynamic Therapy*, vol. 29, p. 101604, Mar. 2020, doi: 10.1016/j.pdpdt.2019.101604.

[70] G. Cennamo, D. Montorio, F. Mirra, C. Comune, A. D'Alessandro, and F. Tranfa, "Study of vessel density in adult-onset foveomacular vitelliform dystrophy with optical coherence tomography angiography," *Photodiagnosis and Photodynamic Therapy*, vol. 30, p. 101702, Jun. 2020, doi: 10.1016/j.pdpdt.2020.101702.

[71] P. Aggarwal, "Machine learning of retinal pathology in optical coherence tomography images," *Journal of Medical Artificial Intelligence*, vol. 2, pp. 20–20, Sep. 2019, doi: 10.21037/jmai.2019.08.01.

[72] D. S. W. Ting *et al.*, "Artificial intelligence and deep learning in ophthalmology," *British Journal of Ophthalmology*, vol. 103, no. 2, pp. 167–175, Feb. 2019, doi: 10.1136/bjophthalmol-2018-313173.

[73] "Convolutional Neural Networks overview." https://towardsdatascience.com/acomprehensive-guide-to-convolutionalneural-networks-the-eli5-way-%0A3bd2b1164a53 (accessed Oct. 01, 2020).

[74] "Convolutional Layers overview." https://machinelearningmastery.com/convol%0Autional-layers-for-deep-learning-neuralnetworks/ (accessed Oct. 01, 2020).

[75] B. Hassan, G. Raja, T. Hassan, and M. Usman Akram, "Structure tensor based automated detection of macular edema and central serous retinopathy using optical coherence tomography images," *Journal of the Optical Society of America A*, vol. 33, no. 4, p. 455, Apr. 2016, doi: 10.1364/JOSAA.33.000455.

[76] B. Nicholson, J. Noble, F. Forooghian, and C. Meyerle, "MAJOR REVIEW Central Serous Chorioretinopathy : Update on Pathophysiology and Treatment," *Survey of Ophthalmology*, vol. 58, no. 2, pp. 103–126, 2013, doi: 10.1016/j.survophthal.2012.07.004.

[77] A. M. Syed, T. Hassan, M. U. Akram, S. Naz, and S. Khalid, "Automated diagnosis of macular edema and central serous retinopathy through robust reconstruction of 3D retinal surfaces," *Computer Methods and Programs in Biomedicine*, vol. 137, pp. 1–10, Dec. 2016, doi: 10.1016/j.cmpb.2016.09.004.

[78] N. F. Mokwa, T. Ristau, P. A. Keane, B. Kirchhof, S. R. Sadda, and S. Liakopoulos, "Grading of Age-Related Macular Degeneration: Comparison between Color Fundus Photography, Fluorescein Angiography, and Spectral Domain Optical Coherence Tomography," *Journal of Ophthalmology*, vol. 2013, pp. 1–6, 2013, doi: 10.1155/2013/385915.

[79] H. S. Sandhu *et al.*, "Automated diabetic retinopathy detection using optical coherence tomography angiography: a pilot study," *British Journal of Ophthalmology*, vol. 102, no. 11, pp. 1564–1569, Nov. 2018, doi: 10.1136/bjophthalmol-2017-311489.

[80] S. Khalid, M. U. Akram, T. Hassan, A. Nasim, and A. Jameel, "Fully Automated Robust System to Detect Retinal Edema, Central Serous Chorioretinopathy, and Age Related Macular Degeneration from Optical Coherence Tomography Images," *BioMed Research International*, 2017, doi: 10.1155/2017/7148245.

[81] T. Hassan, M. U. Akram, B. Hassan, A. Nasim, and S. A. Bazaz, "Review of OCT and fundus images for detection of Macular Edema," in *IST 2015 - 2015 IEEE International Conference on Imaging Systems and Techniques, Proceedings*, 2015, doi: 10.1109/IST.2015.7294517.

[82] S. L. Y. J, C. Madhavi, and M. A. Kumar, "Visual outcome of central serous retinopathy," vol. 3, no. 8, pp. 1885–1888, 2015.

[83] D. Xiang, G. Chen, F. Shi, W. Zhu, and X. Chen,

"Automatic Retinal Layer Segmentation Based on Live Wire for Central Serous Retinopathy," 2017, pp. 118–125.

[84] B. Hassan, R. Ahmed, B. Li, A. Noor, and Z. ul Hassan, "A comprehensive study capturing vision loss burden in Pakistan (1990-2025): Findings from the Global Burden of Disease (GBD) 2017 study," *PLoS ONE*, 2019, doi: 10.1371/JOURNAL.PONE.0216492.

[85] S. H. Freiesleben, J. Soelberg, N. T. Nyberg, and A. K. Jäger, "Determination of the wound healing potentials of medicinal plants historically used in Ghana," *Evidence-based Complementary and Alternative Medicine*, 2017, doi: 10.1155/2017/9480791.

[86] T. Hassan, M. Usman Akram, B. Hassan, A. M. Syed, and S. A. Bazaz, "Automated segmentation of subretinal layers for the detection of macular edema," *Applied Optics*, vol. 55, no. 3, p. 454, Jan. 2016, doi: 10.1364/AO.55.000454.

[87] Z. J. B, Q. C. B, M. Wu, S. Niu, and W. Fan, *Beyond Retinal Layers : A Large Blob Detection for Subretinal Fluid Segmentation in SD-OCT Images*, vol. 1. Springer International Publishing.

[88] Y. Zheng, J. Sahni, C. Campa, A. N. Stangos, A. Raj, and S. P. Harding, "Computerized Assessment of Intraretinal and Subretinal Fluid Regions in Spectral-Domain Optical Coherence Tomography Images of the Retina," *American Journal of Ophthalmology*, vol. 155, no. 2, pp. 277-286.e1, Feb. 2013, doi: 10.1016/j.ajo.2012.07.030.

[89] T. Wang *et al.*, "Label propagation and higher-order constraint-based segmentation of fluid-associated regions in retinal SD-OCT images," *Information Sciences*, vol. 358–359, pp. 92–111, Sep. 2016, doi: 10.1016/j.ins.2016.04.017.

[90] A. Lang *et al.*, "Automatic segmentation of microcystic macular edema in OCT," *Biomedical Optics Express*, vol. 6, no. 1, p. 155, Jan. 2015, doi: 10.1364/BOE.6.000155.

[91] Y. Xu *et al.*, "Dual-stage deep learning framework for pigment epithelium detachment segmentation in polypoidal choroidal vasculopathy," *Biomedical Optics Express*, vol. 8, no. 9, p. 4061, Sep. 2017, doi: 10.1364/BOE.8.004061.

[92] J. De Fauw *et al.*, "Clinically applicable deep learning for diagnosis and referral in retinal disease," *Nature Medicine*, vol. 24, no. 9, pp. 1342–1350, Sep. 2018, doi: 10.1038/s41591-018-0107-6.

[93] K. Gao *et al.*, "Computer Methods and Programs in Biomedicine Double-branched and area-constraint fully convolutional networks for automated serous retinal detachment segmentation in SD-OCT images," vol. 176, pp. 69–80, 2019, doi: 10.1016/j.cmpb.2019.04.027.

[94] A. L. M. Ontuoro, S. E. M. W. Aldstein, B. I. S. G. Erendas, and U. R. S. C. Rfurth, "Joint retinal layer and fluid segmentation in OCT scans of eyes with severe macular edema using unsupervised representation and auto-context," vol. 8, no. 3, pp. 182–190, 2017.

[95] B. Hassan and T. Hassan, "Fully automated detection, grading and 3D modeling of maculopathy from OCT volumes," in *2019 2nd International Conference on Communication, Computing and Digital systems (C-CODE)*, Mar. 2019, pp. 252–257, doi: 10.1109/C-CODE.2019.8680996.

[96] R. V. Teja *et al.*, "Classification and Quantification of Retinal Cysts in OCT B-Scans: Efficacy of Machine Learning Methods," in *Proceedings of the Annual International Conference of the IEEE Engineering in Medicine and Biology Society, EMBS*, 2019, doi: 10.1109/EMBC.2019.8857075.

[97] L.-C. Chen, G. Papandreou, F. Schroff, and H. Adam, "Rethinking Atrous Convolution for Semantic Image Segmentation," Jun. 2017, [Online]. Available: http://arxiv.org/abs/1706.05587.

[98] Y. Zhen *et al.*, "Assessment of central serous chorioretinopathy (CSC) depicted on color fundus photographs using deep Learning," Jan. 2019, [Online]. Available: http://arxiv.org/abs/1901.04540.

[99] Y. Ruan *et al.*, "Multi-phase level set algorithm based on fully convolutional networks (FCN-MLS) for retinal layer segmentation in SD-OCT images with central serous chorioretinopathy (CSC)," *Biomedical Optics Express*, vol. 10, no. 8, p. 3987, Aug. 2019, doi: 10.1364/BOE.10.003987.

[100] K. McDonough, I. Kolmanovsky, and I. V. Glybina, "A neural network approach to retinal layer boundary identification from optical coherence tomography images," in *2015 IEEE Conference on Computational Intelligence in Bioinformatics and Computational Biology (CIBCB)*, Aug. 2015, pp. 1–8, doi: 10.1109/CIBCB.2015.7300299.

[101] T. J. Narendra Rao, G. N. Girish, A. R. Kothari, and J. Rajan, "Deep Learning Based Sub-Retinal Fluid Segmentation in Central Serous Chorioretinopathy Optical Coherence Tomography Scans," in *2019 41st Annual International Conference of the IEEE Engineering in Medicine and Biology Society (EMBC)*, Jul. 2019, pp. 978–981, doi: 10.1109/EMBC.2019.8857105.

[102] J. Long, E. Shelhamer, and T. Darrell, "Fully Convolutional Networks for Semantic Segmentation." in *Proceedings of the IEEE conference on computer vision and pattern recognition*. Oct. 2015, pp. 3431-3440, doi:

10.1109/CVPR.2015.7298965.

[103] Pawan, S. J., *et al.* "Capsule Network–based architectures for the segmentation of sub-retinal serous fluid in optical coherence tomography images of central serous chorioretinopathy." *Medical & Biological Engineering & Computing* 59.6 (2021): 1245-1259, May 2021, doi: 10.1007/s11517-021-02364-4

[104] Chen, Menglu, *et al.* "Automatic detection of leakage point in central serous chorioretinopathy of fundus fluorescein angiography based on time sequence deep learning." *Graefe's Archive for Clinical and Experimental Ophthalmology* (2021): 1-11, Mar 2021, doi: 10.1007/s00417-021-05151-x.